\documentclass[11pt]{article}

\usepackage[verbose=true,letterpaper]{geometry}
\AtBeginDocument{
  \newgeometry{
    textheight=9in,
    textwidth=6.5in,
    top=1in,
    headheight=14pt,
    headsep=25pt,
    footskip=30pt
  }
}
\usepackage{setspace} 
\usepackage{graphicx}
\usepackage{amsmath}
\usepackage{cite}
\usepackage{hyperref}       % hyperlinks
\usepackage{url}            % simple URL typesetting
\usepackage{booktabs}       % professional-quality tables
\usepackage{amsfonts}       % blackboard math symbols
\usepackage{nicefrac}       % compact symbols for 1/2, etc.
\usepackage{microtype}      % microtypography
\usepackage{lipsum}
\usepackage{color}
\makeatletter
\usepackage[version=4]{mhchem}
\renewcommand\@cite[2]{\textsuperscript{#1}}
\makeatother

\usepackage[hang,flushmargin]{footmisc}
\usepackage{multibib}
\newcites{supp}{References}

\begin{document}

\parindent 0em

{\Large\textbf{
Molecular mechanism of ice nucleation on feldspar
}}

\vspace{1cm}

{ \large
	Wanqi Zhou$^{1}$, and
    Pablo M. Piaggi\footnote{CIC nanoGUNE, Tolosa Hiribidea 76, Donostia 20018, San Sebastian, Spain.}$^,$\footnote{Ikerbasque, Basque Foundation for Science, Bilbao 48013, Spain.\\ $^\ast$Corresponding author. Email: pm.piaggi@nanogune.eu}$^{,{\ast}}$
}

\date{\today}% It is always \today, today,

%\begin{abstract}
\section*{Abstract}
Understanding how water transforms into ice at complex surfaces is central to a wide range of natural and technological processes, yet molecular simulations of this transformation have largely been restricted to idealized surfaces or used models with limited predictive power.
Here, we employ advanced molecular simulation tools to study ice nucleation on feldspar, the most abundant mineral in Earth's crust and one of the main ice nucleating particles in the atmosphere.
We develop a machine learning interatomic potential trained on ab initio electronic structure calculations, achieving quantum accuracy across a broad range of feldspar–water interfaces.
Molecular simulations driven by this potential reveal that the feldspar (110) surface uniquely templates interfacial water into an arrangement resembling the structure of water over ice.
Combining this approach with enhanced-sampling and seeded methods, we directly observe nucleation of cubic ice at the (110) feldspar surface, characterize the critical nucleus, and demonstrate that its orientation relative to the mineral surface is consistent with experimental observations. 
On this basis, we identify the (110) surface as the dominant active site, overturning the currently accepted mechanism, which attributes feldspar's exceptional ice nucleation ability to the (100) surface exposed at defects.
These results provide new insight into a principal pathway for atmospheric ice formation. More broadly, they demonstrate the power of ab initio machine learning simulation for the in silico prediction of ice nucleation at surfaces, and clarify the connection between interfacial water structure and the ice-nucleation ability of realistic surfaces.

%\end{abstract}


%\tableofcontents

\newpage
\parindent 1em

%\section*{Introduction}
\label{sec:introduction}

The transformation of liquid water into ice is one of the most emblematic phase transformations, with important implications for fields ranging from cryopreservation and artificial snow production to weather modification and climate science\cite{debenedetti2020metastable}.
The initial stage of this transformation, called ice nucleation, can take place via two distinct pathways.
The first is homogeneous ice nucleation, which occurs in bulk supercooled water at sufficiently low temperatures, namely, below about -38°C.
More commonly, ice forms via heterogeneous ice nucleation on foreign particles, which can occur at much warmer conditions, with some particles forming ice at temperatures as high as -2°C\cite{maki1974ice}.
%As a result, heterogeneous ice nucleation can occur at warmer conditions, with some particles forming ice at temperatures as high as -2°C\cite{maki1974ice}.
%Ice nucleation is the initial stage of this transformation.
%In natural environments, ice formation predominantly occurs via heterogeneous nucleation on foreign particles, which can occur at much warmer conditions, with some particles forming ice at temperatures as high as -2°C\cite{maki1974ice}.
This behavior is known to originate from the ability of such particles to provide sites or surfaces that facilitate ice formation by lowering the free-energy barrier for nucleation\cite{knopf2023atmospheric,kanji2017overview,kalikmanov2012classical}.
However, the molecular mechanisms governing this transformation remain poorly understood, in part due to the difficulty in performing experiments capable of observing ice crystallization at specific interfaces with atomic resolution\cite{huang2023tracking,franceschi2023water,hong2024imaging,dickbreder2024atomic,wang2025molecularly}.

Molecular simulations can provide atomic-scale insight into nucleation processes, and have already played a central role in developing an atomistic picture of heterogeneous ice nucleation.
%and identifying the surface characteristics that govern nucleation efficiency.
While early conceptual models identified lattice match between the structure of ice and that of the substrate as an important surface characteristic governing nucleation efficiency\cite{vonnegut1947nucleation}, subsequent molecular simulation studies proposed alternative characteristics, including the ability of surfaces to induce layering in interfacial water\cite{lupi2014heterogeneous}, the strength of hydrophobic interactions\cite{fitzner2015many}, and the in-plane structure of the first adsorbed water layer as key determinants of a surface's ice-nucleating ability\cite{fitzner2020predicting}.
However, many of these insights have been obtained using highly simplified models for the interatomic forces, such as pair potentials for modeling water-substrate interactions\cite{fitzner2015many,fitzner2020predicting}.
Although some studies have employed sophisticated empirical force fields, they are parameterized using a small number of properties and therefore their ability to describe complex interactions at surfaces is limited\cite{sosso2016microscopic,yu2025unconventional,soni2021unraveling}.
Recently, it has become possible to train machine learning models for the interatomic interactions, often called machine learning interatomic potentials (MLIP), using large datasets of energies and forces computed using ab initio electronic-structure calculations\cite{behler2007generalized,piaggi2024first}.
These models can reproduce the underlying ab initio potential energy surface with high fidelity, enabling consistent accuracy across bulk, defected, and interfacial environments, while naturally capturing key physical effects at interfaces, including many-body interactions, polarization, and short-range electrostatics\cite{piaggi2022homogeneous}.
Despite these advantages, there are currently no reported applications of this approach to understand ice formation at  surfaces.

\begin{figure}[!htbp]
  \centering
  \includegraphics[width=0.8\textwidth]{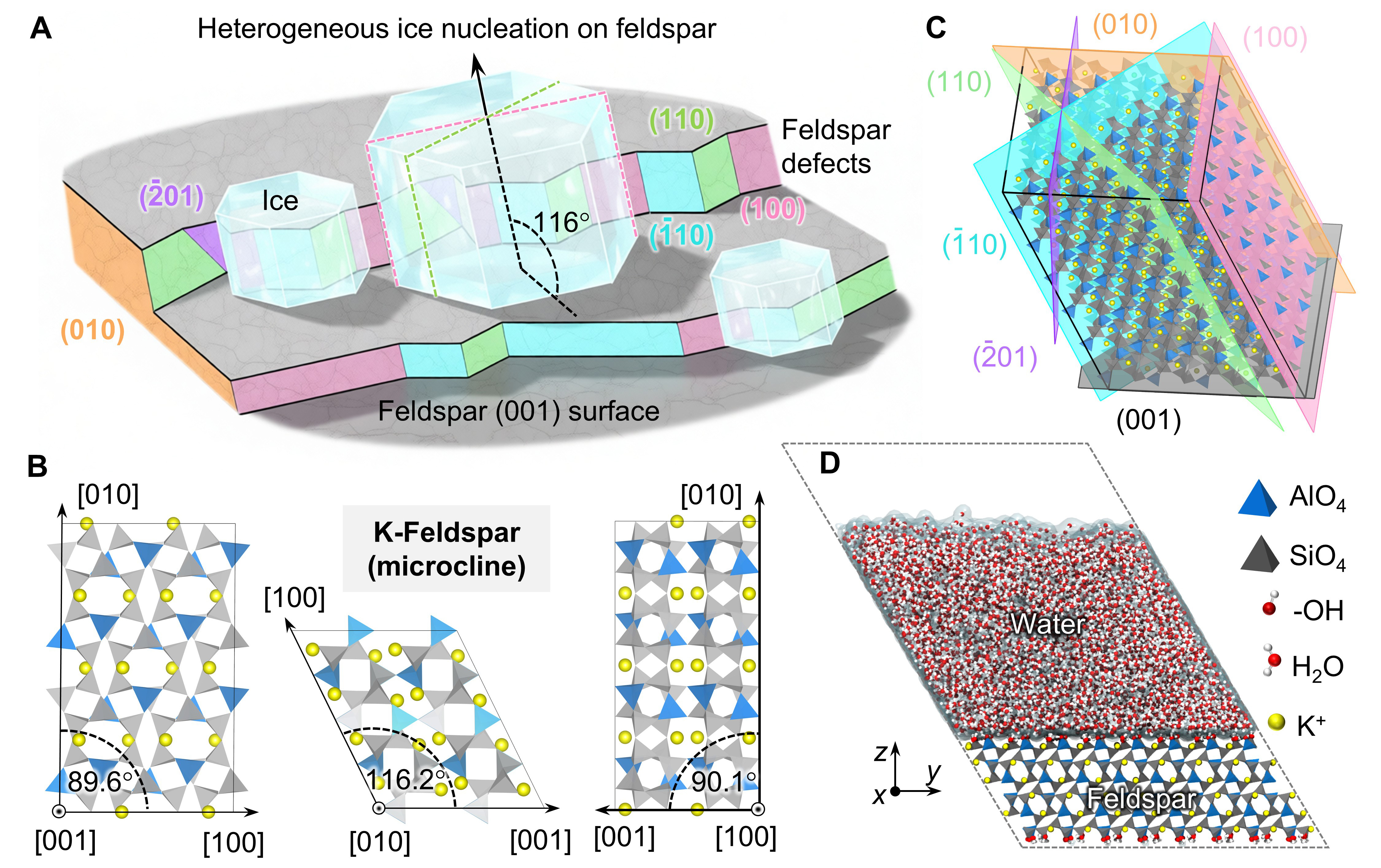}
  \caption{\footnotesize\textbf{Feldspar serves as an effective ice-nucleating particle in the atmosphere owing to structural defects that promote ice nucleation.}  \textbf{(A)} Schematic illustration of the heterogeneous ice nucleation at feldspar defects with exposed non-perfect cleavage planes. The pink and green dashed lines highlight the first and second prismatic surfaces of ice, which are approximately parallel to the (100) and (110) planes of feldspar, respectively. \textbf{(B)} The triclinic crystal structure of microcline K-feldspar. \textbf{(C)} Crystallographic planes of feldspar: (001), (010), (110), (100), ($\bar{1}$10), and ($\bar{2}$01). \textbf{(D)} Configuration for molecular dynamics simulations with one termination of the feldspar (110) surface in contact with water above. }
  \label{fig1}
\end{figure}

Here, we apply this technique to study ice nucleation on feldspar, which is the most abundant mineral in Earth's crust\cite{smith2013feldspar} and the dominant mineral ice-nucleating particle in the atmosphere\cite{atkinson2013importance,yakobi2013feldspar}.
The exceptional ice nucleation ability of feldspar has garnered substantial research interest over the last decade, driving extensive investigations aimed at elucidating the mechanisms of ice nucleation and identifying the nature of the active sites responsible for it\cite{kiselev2017active,whale2017role,holden2019high,soni2019simulations,pach2019pores,kumar2021molecular,keinert2022mechanism,franceschi2023water,dickbreder2024atomic}.
Direct observations using scanning electron microscopy revealed that defects such as steps, cracks, and cavities are responsible for the high nucleation efficiency of feldspar particles\cite{kiselev2017active,pach2019pores,holden2019high}. These studies also found a systematic and reproducible orientation relationship between ice crystals and the feldspar substrate, indicating that ice nucleation is governed by specific crystallographic surfaces exposed at defects\cite{kiselev2017active}. As illustrated schematically in Fig.~\ref{fig1}A, defects such as steps expose patches of distinct crystallographic planes that can act as active sites for ice nucleation.
The currently accepted atomistic mechanism draws its evidence from energy minimization molecular simulations, which suggest that the (100) feldspar surface can promote the formation of hexagonal ice, with its primary prismatic plane aligned with the feldspar surface\cite{kiselev2017active}.
%The currently accepted mechanism is based on energy minimization molecular simulations, which suggest that the primary prismatic plane of hexagonal ice forms on the feldspar (100) surface \cite{kiselev2017active}.
%However, extensive molecular dynamics (MD) simulations by Soni and Patey did not find evidence of high ice nucleation efficiency at the (100) feldspar surface\cite{soni2019simulations}.
However, subsequent extensive molecular dynamics (MD) simulations did not find evidence of high ice nucleation efficiency at the (100) feldspar surface\cite{soni2019simulations}.
Thus, the origin of the exceptional ice-nucleation ability of feldspar remains unresolved.

In this work, we train an MLIP based on density-functional theory calculations, which is able to provide ab initio accuracy for the transformation of water into ice at a broad range of feldspar surfaces exposed to water.
%, including the (001), (010), (110), (100), ($\bar{1}$10), and ($\bar{2}$01) surfaces.
%The interatomic forces driving these simulations are derived from an MLIP trained to achieve a very high accuracy using quantum-mechanical electronic-structure theory calculations.
Simulations driven by this model reveal that the feldspar (110) surface structures interfacial liquid water into an arrangement closely resembling that of water on the surface of ice.
%We clarify the role of surfaces in structuring  interfacial water in resemblance to the structure of water over ice and show that this is a key determinant of their ice nucleation efficiency.
In contrast, none of the other feldspar surfaces show signatures of high ice nucleation potency.
%on the (110) surface of cubic ice (equivalent to the secondary prismatic surface of hexagonal ice).
We also carried out enhanced sampling molecular simulations in order to bring ice nucleation, which is a rare event in the timescale of standard MD simulations, within reach.
The ice clusters that nucleate at the (110) surface during these simulations exhibit a cubic-ice structure, with an orientation fully compatible with the experiment.
Furthermore, we used seeded simulations to characterize the thermodynamics and kinetics of ice nucleation, and find an heterogeneous nucleation temperature in good agreement with experiment.
These results overturn the currently accepted mechanism for this process and shed new light on how realistic surfaces promote ice nucleation.
%Our results shed new light on the molecular mechanism for heterogeneous ice nucleation on K-feldspar, which is a key route for ice formation in the atmosphere.

%\section*{Results and Discussion}
\label{sec:results}
\subsection*{Structure of water at multiple feldspar surfaces}

K-feldspar (KAlSi$_3$O$_8$) is a tectosilicate mineral characterized by a three-dimensional framework of corner-sharing SiO$_4$ and AlO$_4$ tetrahedra, with K$^+$ occupying interstitial sites.
We studied microcline, which is the most stable polymorph of K-feldspar at room temperature and has a triclinic crystal structure (Fig.~\ref{fig1}B).
Microcline shows perfect cleavage along the (001) plane and good cleavage along the (010) plane.
The (110) surface is relatively easy to cleave and ranks just below the (001) and (010) surfaces in terms of cleavage perfection.
Less frequently, it is also known to expose the (100), ($\bar{1}$10), and ($\bar{2}$01) planes\cite{smith2013feldspar}.
These planes are shown in Fig.~\ref{fig1}C.
We systematically explored candidate surfaces and identified 13 distinct surface terminations, labeled by their Miller indices and, where multiple terminations exist for a given surface, by Greek letters.
Further details about the methodology and a description of the surface terminations is provided in the Methods section and in Tables~S1 and S2 of the Supplementary Material (SM).
%%149 words
We then constructed large interfacial configurations (around 25,000 atoms) for all terminations of feldspar in contact with water to investigate the behavior of the candidate surfaces using MD simulations. 
A representative configuration of the (110)-$\alpha$ surface is shown in Fig.~\ref{fig1}D. 
In order to describe the interactions in this system with first-principles accuracy, we trained an MLIP using quantum-mechanical electronic structure calculations for all surface terminations discussed above (see Methods for details), and we used this MLIP to drive MD simulations.
The electronic-structure calculations were based on the strongly constrained and appropriately normed (SCAN) density functional, which was recently employed to study homogeneous ice nucleation and it was found that nucleation rates are predicted in good agreement with experiment\cite{piaggi2022homogeneous}.
%One of us recently leveraged this technique to study homogeneous ice nucleation and found that nucleation rates are predicted in good agreement with experiment\cite{piaggi2022homogeneous}.

\begin{figure}[htbp]
  \centering
  \includegraphics[width=1.0\textwidth]{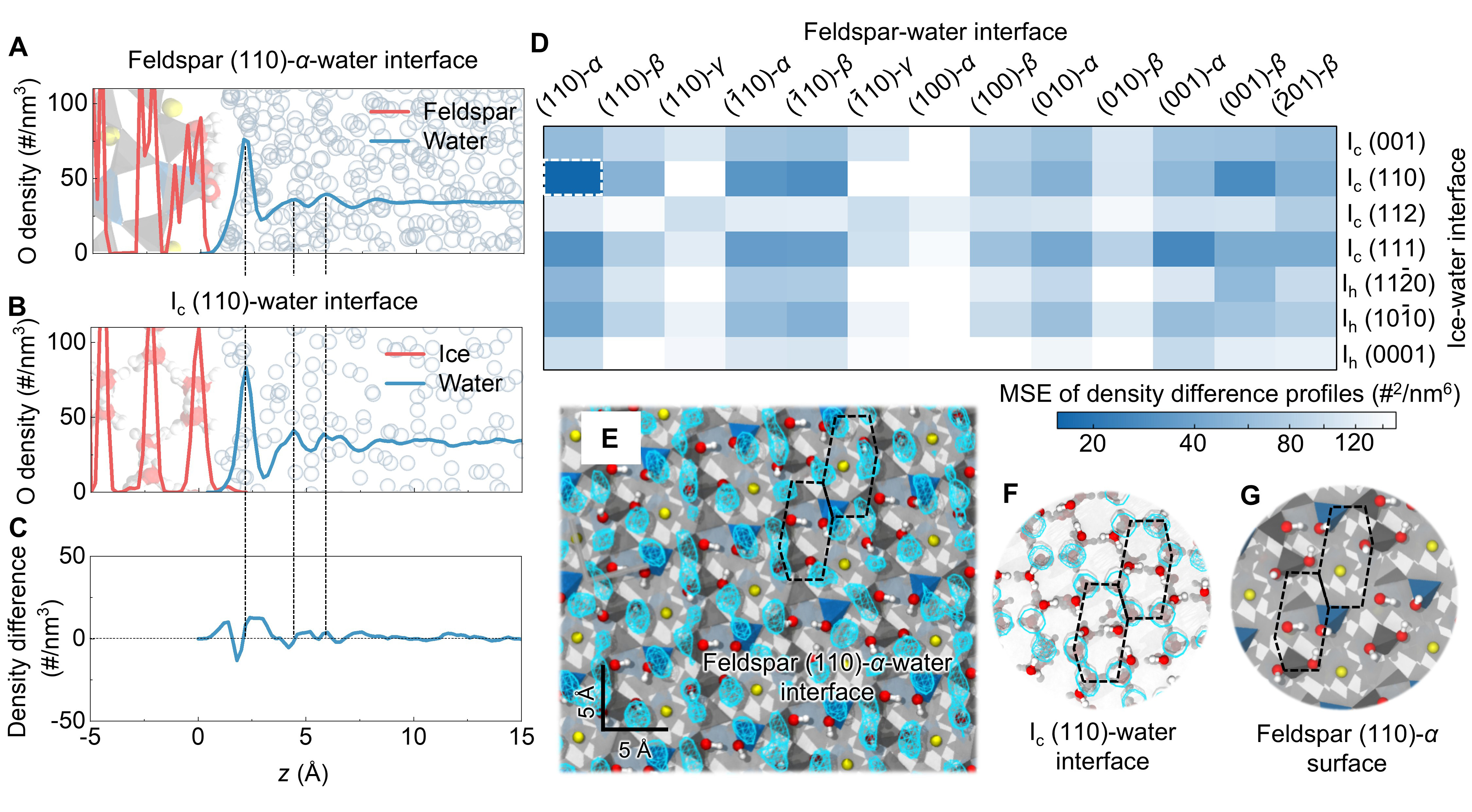}
  \caption{\footnotesize\textbf{Comparison of the water distribution at feldspar–water interfaces and ice–water interfaces.} \textbf{(A, B)} Oxygen, O, density profiles perpendicular to the feldspar (110)-$\alpha$ and cubic ice I$_{\mathrm{c}}$ (110) interfaces with water, respectively,  computed at 18~K of supercooling. \textbf{(C)} Density difference of O between water on the feldspar (110)-$\alpha$ surface and that on the I$_{\mathrm{c}}$ (110) surface. \textbf{(D)} Mean squared error (MSE) between water density profiles at the feldspar–water and ice–water interfaces in the region $z < 15~\text{\AA}$. The 13 feldspar surfaces are labeled according to their Miller indices and different terminations are denoted by Greek letters.
  The seven ice surfaces are labeled according to the polymorph (hexagonal ice I$_{\mathrm{h}}$ or ice I$_{\mathrm{c}}$) and the Miller indices. The white rectangle highlights the location of the lowest MSE. \textbf{(E, F)} Water-density isosurfaces at 70~\text{molecules/nm$^{3}$} in the $x$–$y$ plane, corresponding to the first peak of the density profiles shown in panels \textbf{A} and \textbf{B}, respectively. \textbf{(G)} The surface of the feldspar (110)-$\alpha$ termination. Black lines highlight the pattern of water distribution in panels \textbf{E–G}. }
  \label{fig2}
\end{figure}

To identify the feldspar surfaces that promote ice nucleation among the 13 candidate terminations, we investigated the structure of interfacial water, which is known to be a useful descriptor of the ice nucleation potency of a surface\cite{soni2024using}.
First, we analyzed the water density along the direction perpendicular to the feldspar surfaces.
In Fig.~\ref{fig2}A, we show the water density profile at one particular surface, namely, the feldspar (110)-$\alpha$ surface. 
It is reasonable to expect that a water density similar to that on ice would facilitate ice nucleation.
For this reason, we compared our results with the water density on the surfaces of ice polymorphs that can form under Earth's atmospheric conditions, namely hexagonal ice (I$_{\mathrm{h}}$) and cubic ice (I$_{\mathrm{c}}$).
As an example, we show in Fig.~\ref{fig2}B the water density profile at the ice I$_{\mathrm{c}}$ (110) surface.
To facilitate a clear comparison, we calculated the difference between the two density profiles, as shown in Fig.~\ref{fig2}C.
A small difference indicates that the water density at a given $z$ position is similar for the feldspar and ice surfaces.
Water density profiles for other feldspar and ice surfaces, as well as their difference, are presented in Fig.~S1 and S2.
Furthermore, to quantitatively evaluate the overall deviation of water densities at the feldspar–water interfaces with respect to the densities at ice–water interfaces, we calculated the mean squared error, $\mathrm{MSE} = \frac{1}{N}\sum_{i=1}^{N} \left[\rho_{\mathrm{felds}}(z_i) - \rho_{\mathrm{ice}}(z_i)\right]^2,$ where $\rho_{\mathrm{felds}}(z_i)$ and $\rho_{\mathrm{ice}}(z_i)$ denote the water densities at the feldspar–water and ice-water interfaces, respectively, at position $z_i$, and $N$ is the total number of sampling points within $z$~${\leq}$~15~{\AA}.
In Fig.~\ref{fig2}D, we show the MSE between water densities at the 13 feldspar surfaces, and at the surfaces of ice I$_{\mathrm{h}}$ and ice I$_{\mathrm{c}}$.
Feldspar surfaces show a variety of behaviors, with some of them being very different from all ice surfaces, while a few of them give rise to water density profiles that closely match the water density on one of the studied ice surfaces.
We find that the minimum MSE corresponds to the feldspar (110)-$\alpha$ surface and the ice I$_{\mathrm{c}}$ (110) surface, indicating that the interfacial water densities at these two surfaces are nearly identical (Fig.~\ref{fig2}A and B).
There are also a few other feldspar surfaces with low MSE, yet the feldspar (110)-$\alpha$ surface stands out as the most promising candidate.
Note that the feldspar (100)-$\alpha$ surface, previously proposed as the active site for ice nucleation\cite{kiselev2017active}, shows a very high MSE with respect to all ice surfaces, and thus the structure of water at this surface does not show signatures of high ice nucleation potency.

\subsection*{Structure of water at the (110) surface}

We now turn to a detailed analysis of the structure of water at the feldspar (110)-$\alpha$ surface.
Below, we will refer to this termination simply as the (110) feldspar surface.
The density profile of water at this surface exhibits several peaks near the interface (see Fig.~\ref{fig2}A).
The first peak is notably high, indicating that water molecules in this region are highly structured, and the height, width, and spacing between other density peaks closely match those of the water density profile on the ice I$_{\mathrm{c}}$ (110) surface (see Fig.~\ref{fig2}B).
Furthermore, we examined the in-plane water-density isosurface corresponding to the first peak.
As shown in Fig.~\ref{fig2}E, water molecules within this first peak tend to reside near hydroxyl groups on the feldspar surface, forming a pattern with an irregular hexagon shape that closely resembles that observed on the ice I$_{\mathrm{c}}$ (110) surface (Fig.~\ref{fig2}F).
This pattern does not match the arrangement of water molecules on any other ice surface.
In particular, the pattern does not resemble the water distribution at the (11$\bar{2}$0) plane of ice I${_\mathrm{h}}$, equivalent to the (110) plane of ice I${_\mathrm{c}}$, suggesting a strong polymorph selectivity (see Fig.~S3).
The arrangement of water molecules at feldspar (110) strongly suggests that ice could readily grow on this surface, with the nascent ice I$_{\mathrm{c}}$ crystal oriented such that its (110) plane is aligned with the feldspar (110) surface. 
Note that the distribution of hydroxyl groups at the feldspar (110) surface, shown in Fig.~\ref{fig2}G, does not fully match the distribution of water molecules at the ice I$_{\mathrm{c}}$ (110) surface.
In spite of the absence of a perfect lattice match, the feldspar (110) surface is still able to create a distribution of interfacial water which closely matches the distribution of water at the  ice I$_{\mathrm{c}}$ (110) surface.
This phenomenon has already been observed in other minerals, such as mica and kaolinite\cite{soni2021unraveling,soni2021microscopic}, and represents a different point of view with respect to the early work on heterogeneous ice nucleation by Vonnegut\cite{vonnegut1947nucleation}, which emphasized the role of lattice match.
We also note that the feldspar (110) surface is one of the three candidate planes that can explain the orientation of ice crystals relative to feldspar observed by Kiselev et al.\cite{kiselev2017active}.
We analyzed the two other possible feldspar planes compatible with the data of Kiselev et al., namely, the (100) and the (010) planes, and  we find, in Figs.~\ref{fig2}D, S1 and S4, that the density profiles of interfacial water and its in-plane distribution does not match the arrangements found on any ice surface.

Taken together, the results described above strongly suggest that the (110) surface of feldspar is responsible for its high ice nucleation potency. Below, we shall investigate this hypothesis in further detail.

\subsection*{Formation of ice at the feldspar (110) surface}

\begin{figure}[!htbp]
  \centering
  \includegraphics[width=0.95\textwidth]{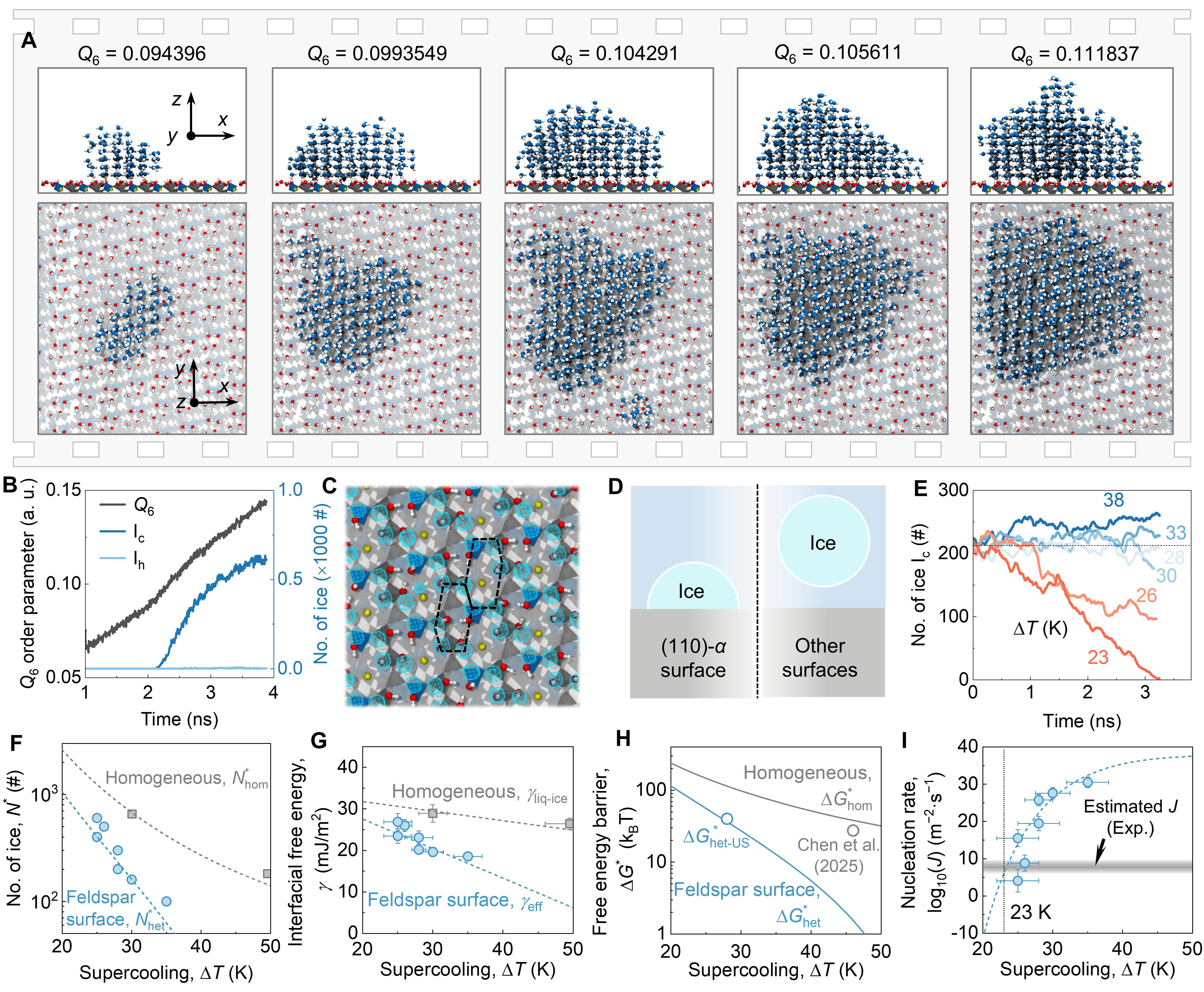}
\caption{\footnotesize\textbf{Formation of ice at the feldspar (110) surface.}  
\textbf{(A)} Snapshots of ice nucleation on the feldspar surface from molecular dynamics simulations with a bias along $Q_6$. Numbers show the $Q_6$ value for each snapshot.
\textbf{(B)} Number of ice-like water molecules and $Q_6$ versus simulation time.  
%\textbf{(C)} Water-density isosurfaces of the first ice layer at 70~\text{molecules/nm$^3$} in the $x$–$y$ plane.  
\textbf{(C)} Water-density isosurfaces of the first ice layer on the feldspar (110) surface at 70~$\mathrm{molecules/nm^{3}}$ in the $x$–$y$ plane.
\textbf{(D)} Schematic illustration of heterogeneous ice nucleation on the feldspar (110)-$\alpha$ surface (left) and homogeneous ice nucleation in bulk water far from other surfaces (right).
\textbf{(E)} Ice-like water molecules over time at different supercooling, ${\Delta}T$.  
\textbf{(F)} Critical cluster size, $N^*$, versus ${\Delta}T$ with a classical nucleation theory (CNT) fit (dashed lines). Data for homogeneous nucleation is taken from Ref.~\citenum{piaggi2022homogeneous}. 
%\textbf{(F)} Interfacial free energy for heterogeneous and homogeneous nucleation versus ${\Delta}T$. The dashed lines represent a fit based on CNT. 
\textbf{(G)} Interfacial free energy, $\gamma$, versus ${\Delta}T$. The dashed lines represent a fit based on CNT. 
\textbf{(H)} Solid lines show the CNT-calculated free-energy barriers for heterogeneous, ${\Delta}G^*_\mathrm{het}$, on the feldspar surface and homogeneous nucleation, ${\Delta}G^*_\mathrm{hom}$. The heterogeneous nucleation data point, ${\Delta}G^*_\mathrm{het-US}$, comes from umbrella sampling (Fig.~S11), and the homogeneous point is from Ref.~\citenum{chen2025exploring}.  
\textbf{(I)} Heterogeneous nucleation rates, $J$, versus ${\Delta}T$ with CNT fit and experimental estimates. 
Further details of the ice-structure identification, CNT calculations, and estimation of nucleation rates are provided in the Methods section.}
%Details of the ice-structure identification (panels \textbf{A}, \textbf{B}), CNT calculations (panels \textbf{E}, \textbf{F}, \textbf{G}, and \textbf{I}) and estimation of nucleation rates in experiments (panel \textbf{I}) are provided in the Methods of the SM.}

  \label{fig3}
\end{figure}

We now seek to observe directly the ice nucleation process on the (110) feldspar surface and elucidate its mechanism.
This cannot be achieved using standard MD simulations due to the high free energy barrier for this process, and thus we resort to the application of enhanced sampling techniques, which bring the nucleation process within affordable simulation time.
In particular, we drive the formation of ice using steered MD simulations with the $Q_6$ Steinhardt order parameter, which is an effective and widely-used collective variable (order parameter) to study crystal nucleation\cite{steinhardt1983bond}. 
This methodology is based on introducing a harmonic bias potential, as a function of the $Q_6$ order parameter, whose center moves from a $Q_6$ value corresponding to liquid water to a value corresponding to ice, thereby promoting crystallization (see details in the Methods section). 

In Fig.~\ref{fig3}A, we show that during the simulation ice nucleates from the interface and eventually forms an approximately hemispherical ice cluster on the surface, as the $Q_6$ value increases under the applied bias potential. 
Figure~\ref{fig3}B shows that the cluster has a structure compatible with ice I$_\mathrm{c}$, instead of ice I$_\mathrm{h}$.
This result is highly robust, with ice I$_\mathrm{c}$ consistently formed in multiple independent simulations regardless of the applied biasing rate (see Fig.~S5).
%Although the introduction of a bias potential could affect polymorph selection, the results of our biased simulations lend additional support to our previous finding that the (110) feldspar surface templates the formation of ice I$_\mathrm{c}$. % Wanqi 06/19 2026
We also investigated the arrangement of water molecules in the interfacial layer and found that it remains largely unchanged before and after ice nucleation.
Once ice is formed, the distribution of water becomes more structured and almost identical to that on the (110) ice I$_\mathrm{c}$ surface (compare Fig.~\ref{fig3}C and Fig.~\ref{fig2}E).
Thus, both our standard (unbiased) and biased simulations of water at the (110) feldspar surface support a strong preference for the formation of ice I$_\mathrm{c}$.
In contrast, simulations of the (100) and other feldspar surfaces using the same bias do not show ice nucleation and growth from the interface and, instead, ice nucleates from the bulk water region (see Fig.~\ref{fig3}D, Fig.~S6 and S7), indicating that these surfaces cannot effectively promote ice nucleation.
Ice formed from bulk water typically consists of a mixture of ice I$_\mathrm{h}$ and ice I$_\mathrm{c}$ (Fig.~S8), suggesting that the preference for ice I$_\mathrm{c}$ formation on the feldspar (110) surface is directly connected to the effect of this surface on the structure of water.
%rather than an artifact induced by external biases.

\subsection*{Thermodynamics and kinetics of ice nucleation}

Next, we investigate the thermodynamics of ice nucleation, including the calculation of the critical cluster size and free energy barrier for this process.
Starting from the ice cluster formed on the feldspar (110) surface and immersed in water, we performed seeding simulations\cite{piaggi2022homogeneous}. 
This methodology is based on carrying out standard MD simulations starting from ice seeds of various sizes at different temperatures (see Fig.~\ref{fig3}E and Fig.~S9).
For each seed size, we study whether the ice cluster grows or melts at a given temperature and use this data to determine the temperature at which a seed is critical, i.e., it has the same probability of growing and melting.
We show in Fig.~\ref{fig3}F the relationship between the temperature and the critical cluster size  ($N^*_\mathrm{het}$), measured as the number of ice-like water molecules.
In Fig.~\ref{fig3}F, we also show the critical cluster sizes in homogeneous nucleation ($N^*_\mathrm{hom}$)\cite{piaggi2022homogeneous}, and it can be observed that for a given supercooling, the critical cluster size at the (110) feldspar surface is between 2 and 4 times smaller than in the homogeneous nucleation case.
Note that the melting temperature in our MLIP is $T_m=308\ \rm K$ (see Fig.~S10), which agrees with previous simulations\cite{piaggi2022homogeneous,piaggi2024first,chen2025exploring} but is higher than the experimental value, and therefore we express all results in terms of supercooling ${\Delta}T=T_m-T$.

With the critical cluster sizes computed above, we can obtain thermodynamic properties using the organizing framework of classical nucleation theory (see Methods section).
In Fig.~\ref{fig3}G, we show the effective interfacial free energy ($\gamma_\mathrm{eff}$) for the heterogeneous nucleation process and the water-ice interfacial free energy ($\gamma_{\mathrm{liq-ice}}$), relevant for homogeneous nucleation, which was reported in ref.~\citenum{piaggi2022homogeneous}.
These two quantities are related by $\gamma_\mathrm{eff}=\gamma_{\mathrm{liq-ice}} f(\theta)$, where $f(\theta)<1$ is a function of the contact angle $\theta$\cite{kalikmanov2012classical}.
In the studied temperature range, $\gamma_\mathrm{eff}$ is lower than $\gamma_{\mathrm{liq-ice}}$ by around 30\%, indicating that the (110) surface indeed is a suitable substrate for ice formation.
Using classical nucleation theory, we also computed the nucleation free energy barrier and the results are shown in Fig.~\ref{fig3}H.
The heterogeneous nucleation free-energy barrier (${\Delta}G^*_\mathrm{het}$) is significantly lower than that of homogeneous ice nucleation (${\Delta}G^*_\mathrm{hom}$), showing the increased likelihood of observing a fluctuation at the surface that reaches the critical cluster size.
We also computed the free energy barrier using umbrella sampling calculations (Fig.~S11) and the results are in good agreement with the estimates from classical nucleation theory.
Using the results described above, we computed the heterogeneous nucleation rate ($J$) and, as we shall see, it provides a direct link with experiment.
In Fig.~\ref{fig3}I, we show $J$ as a function of temperature and we compare it with an estimate of the nucleation rates based on the number of nuclei per unit area in microscopy experiments\cite{kiselev2017active}.
For such nucleation rates, our curve predicts an heterogeneous nucleation temperature for the (110) feldspar surface of 23 K, in very good agreement with the estimates of Atkinson et al. based on the droplet-freezing technique\cite{atkinson2013importance}.
This temperature is characteristic of the active sites that lead to the typical ice nucleation behavior of feldspar, which are the sites studied by Kiselev et al.\cite{kiselev2017active}. We note that other experiments report much higher nucleation temperatures of around -2°C for extremely rare yet highly active sites, which due to their low abundance, have a limited relevance for ice nucleation at atmospheric conditions\cite{harrison2016not}.

\subsection*{Orientation relationship between ice and feldspar}

\begin{figure}[h]
  \centering
  \includegraphics[width=1.0\textwidth]{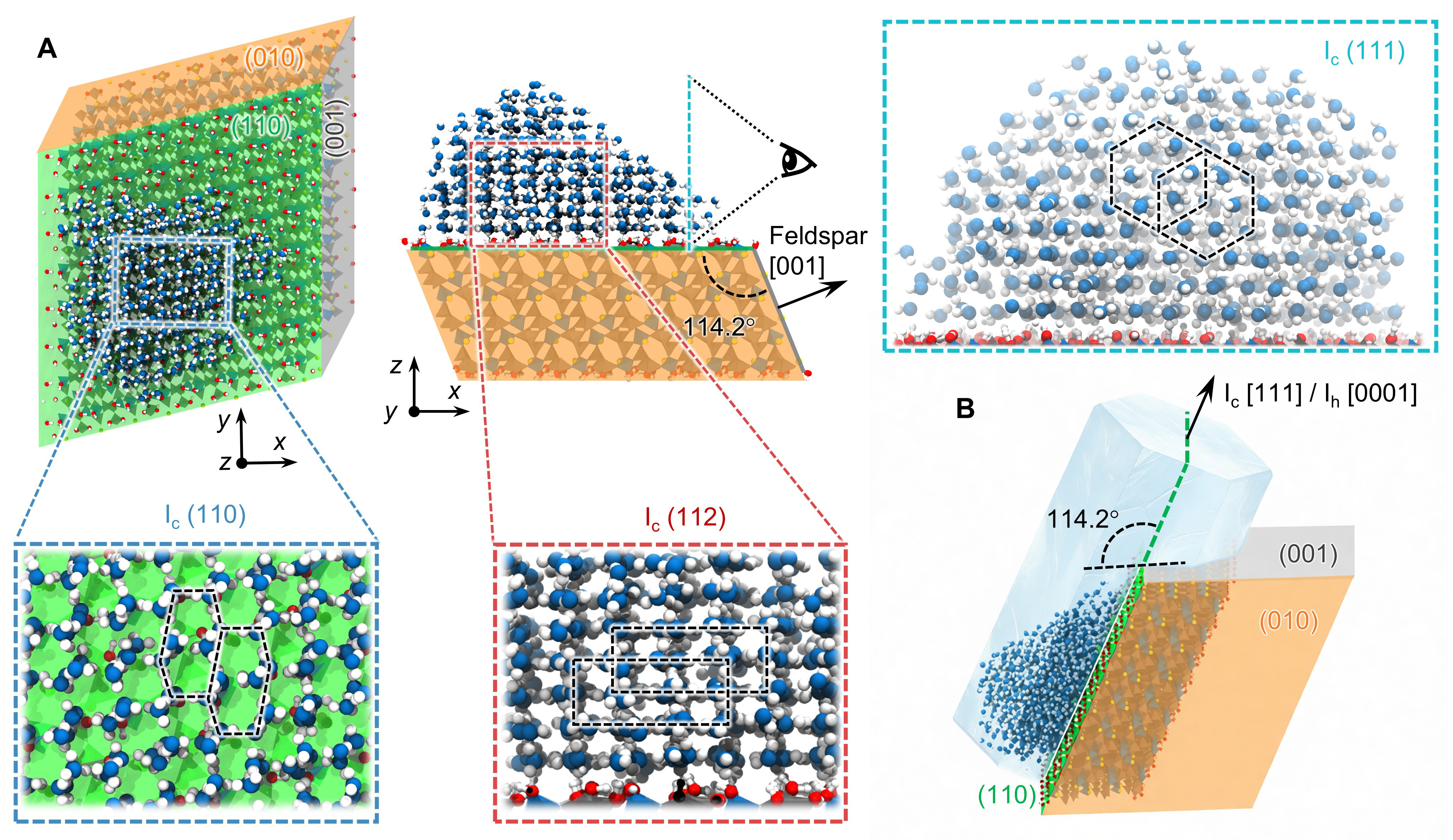}
  \caption{\footnotesize\textbf{Orientation relationship between feldspar and the ice cluster formed in our simulations.} \textbf{(A)} Orientation and lattice of ice formed on the feldspar (110) surface. Visualizations of the ice structure along the $x$, $y$, and $z$ axis are shown. Characteristic patterns on each ice plane are depicted using black dashed lines. \textbf{(B)} Illustration of the ice basal-plane axis orientation relative to the feldspar (001), (010), and (110) planes. Note the basal (0001) plane of hexagonal ice (I$_\mathrm{h}$) is equivalent to the (111) plane of cubic ice (I$_\mathrm{c}$).}
  \label{fig4}
\end{figure}

From the simulations reported above, we can analyze the orientation of ice formed on the feldspar (110) surface and compare the results with experimental observations. 
By direct visual inspection, we confirmed that the  (110) plane of ice I$_\mathrm{c}$ is parallel to the feldspar (110) surface (see Fig.~\ref{fig4}A), as we had already inferred from the results in Fig.~\ref{fig2}.
This result is fully compatible with the experimental data of Kiselev et al.\cite{kiselev2017active}, which can be explained by the (11$\bar{2}$0) plane of ice I$_\mathrm{h}$, equivalent to the ice I$_\mathrm{c}$ (110) plane (see the structural equivalence in Fig.~S12), aligned with the (110) feldspar surface.
%Furthermore, we show in Fig.~\ref{fig4}A that the ice I$_\mathrm{c}$ (111) plane is aligned with the $y-z$ plane (perpendicular to the $x$ axis).
Using the information in Fig.~\ref{fig4}A, we computed an angle of $114.2^\circ$  between the ice I$_\mathrm{c}$ (111) plane and the feldspar (001) plane, as illustrated in Fig.~\ref{fig4}B.
This result agrees very well with the measurements of Kiselev et al.~that found an angle of approximately $116^\circ$ between the basal axis of ice I$_\mathrm{h}$ (corresponding to the (111) plane of ice I$_\mathrm{c}$) and the feldspar (001) surface\cite{kiselev2017active}.
Therefore, the orientation of ice found in our simulation is fully compatible with all the available experimental evidence.
A more thorough analysis of the orientation of ice I$_\mathrm{c}$ with respect to feldspar is presented in Fig.~S13.
We note that experiments consistently show the formation of ice crystals with hexagonal shape, which are the hallmark of ice I$_\mathrm{h}$, while our simulations predict the formation of a cluster of ice I$_\mathrm{c}$ at this surface.
This apparent discrepancy is resolved by noting that microscopic ice clusters may consist of cubic or stacking-faulted ice, but as they grow towards macroscopic sizes they must transform to ice I$_\mathrm{h}$, which is the equilibrium polymorphic form.

\section*{Conclusions}
\label{sec:conclusion}

Our study based on quantum-accurate machine-learning-driven simulations identifies the feldspar (110) surface as the dominant active site responsible for the exceptional ice-nucleation ability of feldspar, overturning the prevailing view that attributes this behavior to the (100) surface.
The results reveal that the ice clusters formed over the (110) feldspar surface are mostly composed of cubic ice, a metastable polymorph with respect to hexagonal ice.
In addition, we show that ice nucleates at the (110) surface with an orientation fully consistent with experimental observations, thereby reconciling atomistic simulations with experimental evidence. 
We encourage future experimental work aimed at directly testing our computational predictions.
More importantly, our simulations uncover the molecular origin of feldspar's remarkable ice-nucleation efficiency.
We show that the (110) surface uniquely structures interfacial water into an arrangement closely resembling that at one of the surfaces of ice, providing an optimal template for nucleation and growth. 
In contrast, other feldspar surfaces that do not induce such ordering show no ice-nucleation activity. 
These findings clarify the role of surfaces in structuring interfacial water in resemblance to the structure of water over ice and show that this is a key determinant of their ice nucleation efficiency.
More broadly, this work demonstrates the power of ab initio machine learning simulations, in combination with enhanced sampling techniques, for predicting the surfaces responsible for ice nucleation and their associated molecular mechanism. 

\section*{Methods}
\label{sec:methods}

\paragraph{Generating feldspar surface terminations}
All possible terminations of the (001), (010), (100), (110), ($\bar{1}$10), and ($\bar{2}$01) planes were generated using the \textsc{SlabGenerator} module of the \textsc{pymatgen} package\cite{tran2016surface}.
A cutoff of 1~\text{\AA} along the surface normal direction was applied to determine whether atoms lie within the same atomic plane, which was used to filter the generated terminations.
There is extensive experimental evidence for the absence of reconstructions in feldspar surfaces and thus we only consider flat unreconstructed surfaces\cite{franceschi2023water, dickbreder2024atomic,fenter2003structure}.
Indeed, no surface reconstruction was observed during our simulations.
Hydrogen atoms were inserted near undercoordinated oxygen atoms at the surfaces, in accordance with recent atomic force microscopy observations, which show full surface hydroxylation\cite{franceschi2023water}.
In addition, surface terminations that did not preserve the integrity of SiO$_4$ and AlO$_4$ polyhedra or failed to maintain charge neutrality were discarded.
Consequently, a total of 13 distinct surface terminations were obtained, including 2 terminations for the (001) plane, 2 for the (010) plane, 2 for the (100) plane, 3 for the (110) plane, 3 for the ($\bar{1}$10) plane, and 1 for the ($\bar{2}$01) plane (see Table~S1 and S2).
Among these, all terminations of the (001), (010), and (100) planes correspond to those previously reported, namely, the (001)-$\alpha$, (001)-$\beta$, (010)-$\alpha$, (010)-$\beta$, (100)-$\alpha$, and (100)-$\beta$ surfaces\cite{soni2019simulations,piaggi2024first}.
Three terminations were identified for the (110) plane, each of them involving the breaking of eight bonds per unit cell (Table~S1).
In one of the terminations, three oxygen atoms of a SiO$_4$ tetrahedron are disconnected from their neighboring two Si and one Al atoms, leaving only one oxygen atom bonded to the surface. 
This configuration is likely unstable and was thus denoted as the (110)-$\gamma$ termination.
In another case, the cleavage requires breaking three bonds of a SiO$_4$ tetrahedron, which is energetically more unfavorable than cases where only one or two bonds are broken. 
This termination was therefore labeled as (110)-$\beta$, while the remaining, more stable termination was designated as (110)-$\alpha$.
The ($\bar{1}$10) plane is a mirror-related counterpart of the (110) plane, and consequently exhibits atomic-scale structural features similar to those of the (110) surface.
Therefore, the terminations of the ($\bar{1}$10) plane were named following the same convention, namely, ($\bar{1}$10)-$\alpha$, ($\bar{1}$10)-$\beta$, and ($\bar{1}$10)-$\gamma$.
The ($\bar{2}$01) plane has only a single termination, which is therefore designated as ($\bar{2}$01)-$\alpha$.
%Building feldspar-water interface models: 

The large feldspar slabs used for molecular dynamics (MD) simulations contain around 25,000 atoms and expose two identical surfaces along the $z$-axis, with atomic-scale structures of both surfaces corresponding to one of the terminations listed above. One of the surfaces is in direct contact with water, forming the water–feldspar interface. A vacuum layer is introduced along the $z$-axis between the water layer and the opposite surface of the feldspar slab to eliminate spurious interactions.

\paragraph{Electronic structure calculations}
Plane-wave DFT calculations were performed using the \textsc{Quantum ESPRESSO} suite for electronic structure calculations v6.4.1\cite{Giannozzi09,Giannozzi17}.
We used the Strongly Constrained and Appropriately Normed (SCAN) exchange and correlation functional\cite{Sun15,Sun16} as implemented in the \textsc{LIBXC} 4.3.4 library\cite{Marques12}.
We employed norm-conserving, scalar-relativistic pseudopotentials\cite{Hamann13} for \ce{K}, \ce{Al}, \ce{Si}, \ce{O}, \ce{H} parametrized using the PBE\cite{Perdew96} functional with 9, 11, 4, 6, and 1 valence electrons, respectively.
Kinetic energy cutoffs of 110 and 440~Ry were used for the wave functions and the charge density.
We used only the $\Gamma$-point in our calculations.
All other parameters were set to their default values in \textsc{Quantum ESPRESSO}\cite{Giannozzi09,Giannozzi17}.

\paragraph{Machine learning interatomic potential}

The smooth-edition of the Deep Potential methodology developed by Zhang et al., as implemented in \textsc{DeePMD-kit} v2.10.0, was used to train the machine-learning interatomic potentials (MLIPs)\cite{zhang2018deep,wang2018deepmd,zhang2018end}.
An active learning strategy was employed during the training process, as shown in Fig.~S14. 
First, MD simulations driven by our previous MLIP (Ref.~\citenum{piaggi2024first}) were performed to generate a series of configurations of the water-feldspar interfaces, covering all 13 feldspar terminations. 
The energies and forces for these configurations were then calculated using SCAN DFT to expand our dataset reported in Ref.~\citenum{piaggi2024first}.
The resulting dataset, which included new configurations and their corresponding energies and atomic forces, was used to train a set of four MLIPs. 
Based on the newly trained MLIPs, additional configurations were explored, and this cycle was repeated iteratively until a high-accuracy MLIP with SCAN-level precision was obtained.
Figures~S15 and S16 show that our MLIPs have excellent and uniform accuracy across all studied terminations, with RMS errors in the energy and forces of 0.498 meV/atom and 95.7 meV/\AA, respectively.
Settings of \textsc{DeePMD-kit} are identical to those used in Ref.~\citenum{piaggi2024first}.

\paragraph{MD simulations}
We performed MD simulations and enhanced sampling simulations using \textsc{LAMMPS}\cite{Plimpton95,thompson2022lammps} interfaced with the \textsc{PLUMED} plugin\cite{tribello2014plumed,bonomi16plumed}.
All simulations employed the custom-trained MLIPs described above to model atomic interactions. 
We used the standard atomic weights for the masses of all elements (\ce{K}, \ce{Al}, \ce{Si}, \ce{O}) except \ce{H}, for which we used a mass of 2~grams/mol, in order to improve the stability of the integration of the equations of motion.
Simulation boxes with periodic boundary conditions in all directions were used.
The time step for the integration of the equations of motion was 0.5~fs.
The simulation models were first energy-minimized using the conjugate gradient (CG) method and then equilibrated using  MD simulations in the $NVT$ ensemble.
A stochastic velocity-rescaling thermostat\cite{Bussi07} with a 0.1~ps relaxation time was used to control the temperature. 
For studies of water structure on multiple surfaces, simulations were performed at 290 K (supercooling ${\Delta}T$=18~K) for more than 3~ns.
During the simulations, the K, Al, and Si atoms were constrained to their initial positions by a harmonic spring with force constant of 20~$\mathrm{eV/\AA^2}$.
The trajectory was saved every 1000 steps.

\parindent 1em
As standard MD simulations cannot capture the nucleation process within affordable simulation times, we employed an enhanced sampling method, namely steered MD simulations guided by the $Q_6$ Steinhardt order parameter\cite{steinhardt1983bond,van1992computer}, as implemented in PLUMED\cite{tribello2014plumed,bonomi16plumed}.
The oxygen atoms of water molecules were used to calculate the $Q_6$ Steinhardt order parameter, which quantifies local structural ordering. A moving restraint with a force constant of $2 \times 10^{6}\ \mathrm{kJ/mol}$ was applied to this collective variable. 
The simulations were performed at 300~K (supercooling ${\Delta}T$=8~K).
The target value of the restraint was gradually increased from 0.05 to 0.2 over a 5~ns simulation.
The order parameter was recorded every 1000 steps to monitor the progress of nucleation throughout the simulation, and the trajectory was saved every 10000 steps.
After the steered MD simulations, configurations were extracted from the trajectory and used as the initial models for seeding simulations without bias.
A very similar steered MD protocol was extensively validated for studying ice-binding surfaces in Ref.~\citenum{naullage2020computationally}.
The simulations reported in that work show that while using biased simulations may result in a different polymorph than in the unbiased simulations, this impacts mostly the ice that forms far from the surface.

The free energy profile of ice nucleation was calculated using the umbrella sampling method. 
The $Q_6$ Steinhardt parameter was used as collective variable, and 41 windows in the range $0.05 \leq$~$Q_6$~$\leq 0.11$ with a width of 0.0015 were used.
In each window, the $Q_6$ parameter was restrained by a harmonic potential with a spring constant of $1 \times 10^{6}\ \mathrm{kJ/mol}$. 
Each simulation was performed for 3~ns. 
The time series of the $Q_6$ parameter for all windows were then used to construct the free energy profile via the weighted histogram analysis method (WHAM), following the implementation of Grossfield\cite{Kumar1992A,Roux1995A,grossfield-wham}. 

Visualization and analysis of the MD trajectories were performed using VMD\cite{Humphrey1996A}. 
To identify ice-like atomic environments we employed the Identify Diamond Structure modifier in OVITO\cite{ovito} using the oxygen atoms.
We quantified the number of ice-like water molecules by counting the number of oxygen atoms classified as having a diamond-like structure, and we ignored their first and second neighbors with imperfect structures, which are sometimes included in this type of calculation.
The results are fully consistent with the values obtained using the Polyhedral Template Matching modifier\cite{larsen2016robust,piaggi2022homogeneous}.
%In the seeding simulations, the number of ice-like water molecules corresponds to the number of ice atoms excluding the first and second neighbors, which is consistent with the values obtained from the Polyhedral Template Matching modifier\cite{larsen2016robust}.

\paragraph{Obtaining thermodynamic properties using classical nucleation theory}

 The effective surface free energy for the heterogeneous nucleation process can be computed from an equation that closely follows its counterpart for homogeneous nucleation, namely,
\begin{equation}
\gamma_\mathrm{eff} = \left( \frac{3 N^*_\mathrm{het}}{32 \pi} \right)^{1/3} \rho_{\rm ice}^{2/3} |\Delta \mu|,
\end{equation}
where $\gamma_\mathrm{eff}$ is an effective surface free energy that combines information about the water-ice surface free energy and the contact angle, $\rho_{\rm ice}$ is the density of ice, and $\Delta \mu$ is the chemical potential difference between liquid water and ice.
$\rho_{\rm ice}$ and $\Delta \mu$ are easily computed and were taken from previous SCAN-MLIP-based simulations reported in ref.~\citenum{piaggi2022homogeneous}.
According to classical nucleation theory, $\gamma_\mathrm{eff}$ is related to the nucleation free energy barrier, $\Delta G^*_\mathrm{het}$, as follows,
\begin{equation}
\Delta G^*_\mathrm{het} = \frac{16 \pi \gamma_{eff}^3}{3 \rho_{\rm ice}^2 |\Delta \mu|^2}.
\end{equation}
Another way to characterize the ice nucleation process is through the heterogeneous nucleation rate, which can be calculated as,
\begin{equation}
J_\mathrm{het} = \rho_{\mathrm{surf}} Z f \exp(-\beta \Delta G^*_\mathrm{het}),
\end{equation}
where $\rho_{\mathrm{surf}}=8.05$ molecules/nm$^2$ is the density of liquid water at the first layer in contact with the substrate's surface, $Z$ is the Zeldovich factor, $f$ is the attachment rate of molecules to the critical nucleus, $\beta = 1/(k_B T)$ with $k_B$ being the Boltzmann constant and $T$ the temperature. $Z$ and $f$ were taken from ref.~\citenum{piaggi2022homogeneous}.

\paragraph{Estimation of nucleation rates in previous experiments}
We estimated the experimental ice nucleation rate using Fig.~2A of Ref.~\citenum{kiselev2017active}.
In this figure, the nucleation sites of individual ice crystals were plotted during eight subsequent nucleation–evaporation cycles on the defects of a feldspar (001) cleavage surface within 100~s. 
Approximately 200 sites were marked over the eight cycles in the observed region. 
The area of the observed (001) cleavage surface is $A \approx 453.5 \times 524.5~\mu\mathrm{m}^2$. 
Clearly, the exposed (110) surface area is much smaller than that of the (001) surface, and thus the area of (110) surface was estimated using a scaling factor, $f$, such that the effective exposed area of (110) surface is $fA$. 
Therefore, the ice nucleation rate, $J_{\rm EXP.}$, can be estimated as
\begin{equation}
J_{\rm EXP.} = \frac{N_{\rm site}}{f A \cdot t} ,  
\end{equation}
where $N_{\rm site}$ represents the average number of nucleation sites per cycle, and $t$ is the duration of each nucleation–evaporation cycle.
The estimated nucleation rate is 
$1 \times 10^6 \, f~\mathrm{m^{-2}\,s^{-1}}$.
The scaling factor $f$ is much smaller than 1, and assuming $f$ to be in the order of $10^{-2}$ to $10^{-4}$, we computed the range 
$ 1 \times 10^6~\mathrm{m^{-2}\,s^{-1}} < J <  1 \times 10^{10}~\mathrm{m^{-2}\,s^{-1}}$, which is highlighted in Fig.~3I.

\bibliographystyle{naturemag}
\bibliography{science_template}% Produces the bibliography via BibTeX.

\section*{Acknowledgements}

We are grateful for the computational resources provided on the Hyperion cluster at the Donostia International Physics Center (DIPC) and on MareNostrum 5 ACC at the Barcelona Supercomputing Center, granted through the Spanish Supercomputing Network (RES) allocations FI-2024-2-0026 and FI-2024-3-0028.
We thank Prof. Hu Qiu from Nanjing University of Aeronautics and Astronautics for valuable suggestions on the data presentation in this work.
W.Z. was funded by the 'María de Maeztu' Units of Excellence program No. CEX2020-001038-M.
P.M.P. acknowledges funding from the Marie Sklodowska-Curie Cofund Programme of European Commission project H2020-MSCA-COFUND-2020-101034228-WOLFRAM2.

\section*{Author contributions:}
W.Z. performed research. W.Z. and P.M.P. designed research, analyzed data, and wrote the manuscript. P.M.P conceived the project.
\section*{Competing interests:}
There are no competing interests to declare.
\section*{Data and materials availability:}
The scripts used for generating slab surfaces, the dataset for training the MLIPs, the input files for MLIP training, the final trained MLIP, as well as the input files for the MD simulations and DFT calculations reported in this work, are openly available on Zenodo (DOI \url{https://doi.org/10.5281/zenodo.17610332})\cite{dataset}.

\end{document}